\documentclass[%
 reprint,
 amsmath,amssymb,
 aps,
]{revtex4-2}

\usepackage{graphicx}% Include figure files
\usepackage{dcolumn}% Align table columns on decimal point
\usepackage{bm}% bold math
\usepackage{tabularx}
\def\be{\begin{equation}}
\def\bea{\begin{eqnarray}}
\def\eea{\end{eqnarray}}
\def\ee{\end{equation}}

\begin{document}

%\preprint{APS/123-QED}

\title{Radiation-Driven Magnetic Fields in Sub-Keplerian Accretion Flows:
An Alternative to MRI}% Force line breaks with \\

\author{Mukesh Kumar Vyas}
% \altaffiliation[]{Bar Ilan University, \\ Ramat Gan, Israel,
 %5290002} 

 \author{Asaf Pe'er}
 \altaffiliation[]{Bar Ilan University, \\ Ramat Gan, Israel,
 5290002} 
 \email{mukeshkvys@gmail.com}%Lines break automatically or can be forced with \\
 %\homepage{http://www.mukeshvyas.net}
%\affiliation{%


%\affiliation{
% \affiliation{
%  Third institution, the second for Charlie Author
% \affiliation{%
%  Authors' institution and/or address\\
%  This line break forced with \textbackslash\textbackslash
% }%

% \collaboration{CLEO Collaboration}%\noaffiliation

\date{\today}% It is always \today, today,

\begin{abstract}
Large scale magnetic fields play a crucial role in shaping the dynamics and structure of the inner parts of accretion disks and the resulting jets from the vicinity of black holes and neutron stars. Currently, the primary mechanism considered for generating and amplifying large scale magnetic fields in accretion disks is the magneto-rotational instability (MRI). Here we show that non-conservative radiation fields provide a rapid and independent mechanism for generating and amplifying magnetic fields in accretion flows. A luminous, compact corona generates a radiation-driven poloidal field that is subsequently amplified by differential rotation, yielding quadratic magnetic-field growth and, over a broad range of coronal luminosities and sizes and MRI amplification locations, magnetization on timescales comparable to or shorter than those of the MRI. The mechanism therefore provides not merely an alternative to MRI, but an additional and, over a substantial region of parameter space, dominant channel for magnetic-field generation from initial conditions with zero magnetic field. Since the radiation-driven source arises directly from the non-conservative nature of the radiation field, such magnetic-field generation is an inevitable consequence of sufficiently strong and anisotropic radiation sources in black-hole accretion flows. More generally, the mechanism requires only a non-conservative radiation field and differential plasma motion, and is therefore expected to operate in a broad range of luminous astrophysical systems, including active galactic nuclei, gamma-ray bursts, and tidal disruption events.
\end{abstract}

\maketitle

%\tableofcontents

\section{Introduction}
\label{sec_intro}

Magnetic fields are a fundamental ingredient in astrophysical sources across different scales. They regulate the shape and dynamics of accretion disks, govern angular momentum transport, play a key role in disc turbulence as well as in jet launching and jet properties, and have a major effect on the emission of radiation from the plasma \citep{1977MNRAS.179..433B,1982MNRAS.199..883B,1991ApJ...376..214B, 1994MNRAS.267..235L, 2010ApJ...713...52D, 2014ARA&A..52..529Y, 2019arXiv190409677J, 2019ARA&A..57..467B}.
%
%1988ASSL..133.....R,2002RvMP...74..775W,2002PhT....55l..40K,2005PPCF...47A.205S,2025arXiv250815532W. 
Observations and numerical simulations indicate the presence of strong, large-scale magnetic fields in the innermost regions of black-hole accretion discs, where relativistic jets and powerful winds originate \citep{1995ApJ...440..742H,2003PASJ...55L..69N, 2011MNRAS.418L..79T, 2020ARA&A..58..407D,2021NewAR..9201610K,2022MNRAS.511.3795N,2023ApJS..264...32B, 2025arXiv250815532W}. Recent observations with the Event Horizon Telescope have revealed strong and organized magnetic fields on horizon scales in M87$^*$ and Sgr~A$^*$, with an ordered spiral/azimuthal polarization structure around the emission ring \citep{event2021first, 2024ApJ...964L..25E}. 
%Despite their importance, the physical origin of magnetic fields in accretion flows remains an open problem.

The standard framework for magnetic-field amplification in accretion discs is the magnetorotational instability (MRI) and the associated dynamo action \citep{1978mfge.book.....M,1991ApJ...376..214B,1995ApJ...440..742H, 1995ApJ...446..741B, 2003ARA&A..41..555B}. MRI provides extremely rapid exponential amplification of a pre-existing magnetic field, with a maximum linear growth rate of order the local orbital frequency. It can therefore efficiently transform a weak seed field into a dynamically important field. %However, MRI is fundamentally an amplification mechanism: it does not generate magnetic flux from an initially unmagnetized state. The strength and geometry of the resulting field can also depend on the available net magnetic flux and on nonlinear dynamo processes \citep{1991ApJ...376..214B,1995ApJ...440..742H,2020ARA&A..58..407D,2022MNRAS.511.3795N}. Thus, even when MRI provides an efficient route to strong magnetic fields, the origin of the seed field remains a distinct physical question.

A physically distinct route to magnetic-field generation arises from non-conservative radiation forces, $\nabla\times{\bf F}_{\rm rad}\neq0$, which generate currents and magnetic fields in the surrounding plasma \citep{1977A&A....59..111B,1998ApJ...508..859C, 2010ApJ...716.1566A,2014ApJ...782..108S}. This mechanism underlies the Poynting--Robertson and cosmic-battery models \citep{1977A&A....59..111B,2006ApJ...652.1451C,2015ApJ...805..105C}. Earlier studies of this mechanism produced weak fields on secular timescales several orders of magnitude longer than the viscous timescale (the characteristic timescale for accreting plasma to drift inward through the disc into the black hole) of accretion flows around black-holes \citep{1977A&A....59..111B,1998ApJ...508..859C,2002ApJ...580..380B,2006ApJ...652.1451C,2008ApJ...674..388C,2019Galax...7...12C,2014ApJ...794...27K,2015ApJ...805..105C}. 

As was recently shown \citep{2025ApJ...988L..59V,2026ApJ..1006..120V}, the inclusion of the $\Omega$-effect (the amplification of poloidal magnetic fields into toroidal fields by differential rotation), combined with the assumption of a compact, rotating corona, results in a radiation-generated field much stronger than previously thought. Taking this effect into account, it was shown that
the radiation-generated field can be amplified to dynamically significant strengths within the viscous timescale of the inner flow. Still, under the conditions considered in these works, MRI remains the fastest and leading amplification mechanism, with growth timescale much shorter than the viscous timescale. The key question therefore remains, whether there exist physical conditions under which radiation-driven generated magnetic field can itself operate on timescales comparable to, or shorter than, MRI.

In this Letter, we show that a luminous, compact radiation source with a
monotonically decreasing flux (as a function of the distance from the central compact object) embedded in a sub-Keplerian rotating flow
can magnetize the surrounding plasma on timescales comparable to or shorter
than MRI. Over a broad region of the coronal luminosity
parameter space, radiation therefore provides an independent and potentially dominant mechanism for the rapid generation and amplification of large-scale magnetic fields in luminous accretion flows. While we take here fiducial values relevant for stellar-size  X-ray binaries, the mechanism has a broad applicability for many systems, including gamma-ray bursts, active galactic nuclei, and tidal disruption events, where strong, non-conservative radiation fields exist and their role in magnetic field generation remains to be explored.
\section{Theoretical Model and Computational Scheme}
\label{sec:model}
The magnetic field $\mathbf{B}$ satisfies the time ($t$) evolution kinematic induction equation
\begin{equation}
\frac{\partial \mathbf{B}}{\partial t}
=
\nabla\times(\mathbf{v}\times\mathbf{B})
-
\nabla\times
\left[
\frac{c}{4\pi e n_e}
(\nabla\times\mathbf{B})\times\mathbf{B}
\right]
+\mathbf{F}_0
+\eta\nabla^2\mathbf{B},
\label{eq:induction}
\end{equation}
where $\mathbf{v}$ is the prescribed velocity of magnetized plasma, $e$ is the magnitude of
the electron charge, $n_e$ is the prescribed electron number density, $c$ is the speed of light, and $\eta$ is the magnetic diffusivity.
The first term on the right-hand side describes ideal magnetic induction and advection, the second is the Hall term, and the last
describes resistive diffusion. The radiation-driven source term depends upon the curl of the radiation force per electron, and in the Thomson regime is given by
\begin{equation}
\mathbf{F}_0
=
-\frac{\sigma_{\rm T}}{e}
\nabla\times\mathcal{F}_{\rm rad},
\label{eq:radiation_source}
\end{equation}
where $\sigma_{\rm T}$ is the Thomson scattering cross section and
$\mathcal{F}_{\rm rad}$ is the radiation flux. Thus, it is a source term and the magnetic field is generated wherever the radiation flux has a non-zero curl. 
\begin{figure*}
    \centering   
    \includegraphics[width=18.0 cm]{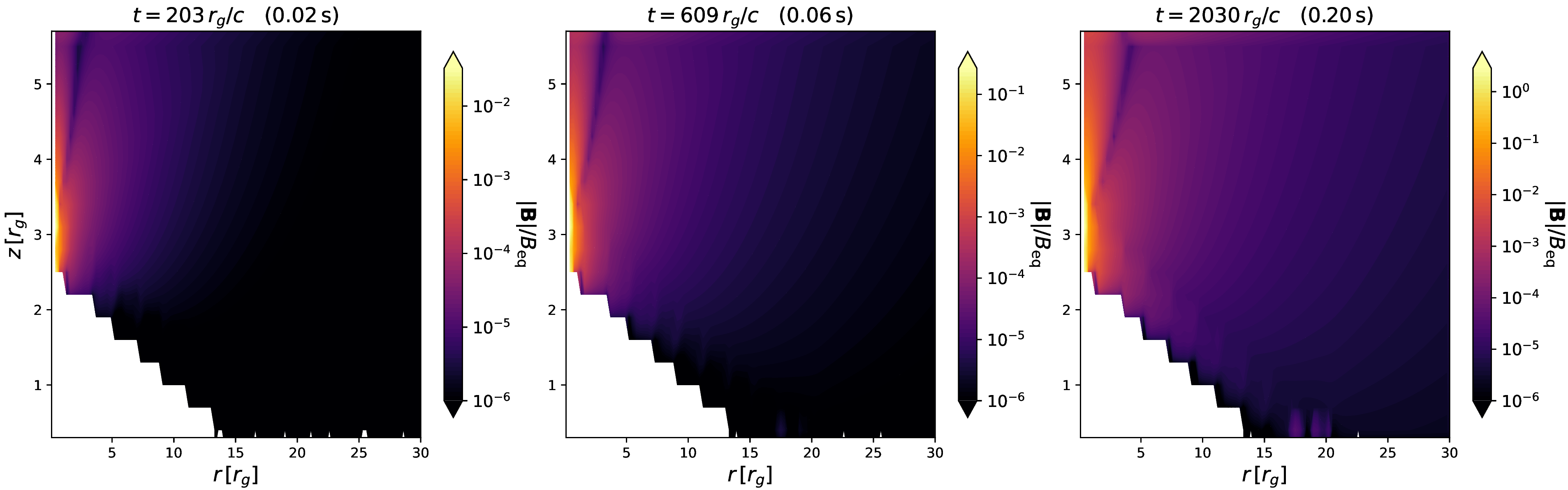}
    \caption{Spatial distribution of the magnetic-field strength normalized
    by the equipartition field, $|\mathbf{B}|/B_{\rm eq}$, at three
    representative times for a corona with $r_c=20 r_g$ and $l_c=1$. The
    logarithmic colour scale shows the rapid growth and spatial development
    of the radiation-generated magnetic field.}
    \label{fig:Bmag_Beq_evolution}
\end{figure*}
This field generation mechanism is universal, and is applicable to many astronomical sources, in which there exists a sufficient non conservative radiation field. As an    application, we consider the kinematic evolution of magnetic fields around a black hole of stellar mass. 
%The radiation field is generated by the prescribed rotating accretion flow, which assumes the form of a Keplerian disc truncated at some inner radius $x_c$ and a luminous corona at radii $x<x_c$. The produced radiation field magnetizes the surrounding plasma. 
%\textbf{The physical
%origin of the radiation-driven source and its detailed construction from the disc--corona radiation field were established in \citet{2025ApJ...988L..59V}; the subsequent coupling of the generated field to rotational induction (inclusion of $\Omega$ effect) was studied in
%\citet{2026ApJ..1006..120V}. We ignore resistive effects and the Hall term in equation \ref{eq:induction} because their physical effects are limited in the current scenario \citep{2026ApJ..1006..120V}.}

Here we consider a physically motivated luminous inner-flow configuration in which the coronal emission is radially stratified and the coronal plasma rotates with sub-Keplerian velocity \footnote{For an extended corona, which may reach tens to hundreds of $r_g$, solid-body rotation is unlikely to be appropriate. Likewise, homogeneous specific intensity is mainly suitable for a very compact corona. We therefore adopt radially varying sub-Keplerian rotation and radius-dependent specific intensity.}. We use cylindrical coordinates $(r,\phi,z)$, where $r$ is the
cylindrical radial coordinate, $\phi$ is the azimuthal coordinate, and
$z$ is the vertical coordinate, and define the dimensionless radius
$x = r/r_g$, with 
$
r_g={2GM}/{c^2},
$
being the Schwarzschild radius, $M$ is the black-hole mass,
and $G$ is the gravitational constant. We consider
$M=10\,M_\odot$, where $M_\odot$ is the solar mass.

We adopt a radially dependent specific intensity, $I_{\rm cor}(x)\propto x^{-3},$
with its normalization fixed by the total coronal luminosity $L_{\rm cor}=l_cL_{\rm Edd}$. The corona is represented by the inclined
inner emitting surface described in Appendix~\ref{app:corona_geometry} having horizontal width $x_c$ and height $z_m$ in Schwarzschild radius, where the underlying Keplerian accretion disc is truncated at its inner radius $x_c$. For cylindrical coordinate $x$, the outer accretion disc (region $x>x_c$) follows the local Keplerian velocity in the
Paczy\'nski--Wiita potential \citep{1980A&A....88...23P}, 
$
v_\phi = v_{\rm K}(x)
=
\left[\frac{x}{2(x-1)^2}\right]^{1/2}c,
$
while the inner corona ($x<x_c$) rotates with sub-Keplerian speed $f_{\rm rot}v_{\rm K}(x)$, with $f_{\rm rot} < 1$. In this letter we keep $f_{\rm rot} =0.5$. 
Inner corona operate as radiation sources and magnetize the plasma existing above the disc corona. 
This magnetic field is continuously amplified by rotational shear (standard $\Omega$ effect). We show that this physically motivated configuration can magnetize the surrounding flow faster than MRI over a broad region of
parameter space, making radiation an independent and potentially dominant channel for magnetic-field generation.

The prescribed axisymmetric magnetized plasma outside and surrounding
the disc-corona system has Keplerian rotaion with no radial and vertical components considered for simplicity, namely $\left\{ v_r, v_z \right\} \ll v_\phi$. % $\hat{\boldsymbol\phi}$ is the corresponding cylindrical unit vector. 
Axisymmetry implies that the quantities are independent of $\phi$. The velocity field is prescribed and does not evolve dynamically.

At a field point $\{r,z\}$, the radiation flux is obtained by integrating
the specific intensity from all visible surface elements of the corona, weighted by their projected solid angle and propagation
direction. In compact form,
\begin{equation}
\mathcal{F}_{\rm rad}(r,z)
=
%\int_{\rm disc} I_{\rm disc}\,\hat{\mathbf n}\,d\Omega+
\int_{\rm cor} I_{\rm cor}\,\hat{\mathbf n}\,d\Omega ,
\label{eq:Frad_integral}
\end{equation}
where $\hat{\mathbf n}$ is the unit vector from each emitting surface
element to the field point and $d\Omega$ is its differential solid
angle. The effects of surface projection, visibility, and the motion of
the emitting material are calculated following 
\citet{2025ApJ...988L..59V,2026ApJ..1006..120V}. 

%\subsection{Numerical evolution}
The magnetic field is initialized with no seed field,
$ \mathbf{B}(t=0)=0, \label{eq:Binitial}$  so that all magnetic flux is generated by the radiation source rather
than supplied as an initial condition. Equation~(\ref{eq:induction}) is
solved assuming axisymmetry for the three components
$\{B_r,B_\phi,B_z\}$ in the $(r,z)$ domain using a Python-based
finite-difference Method-of-Lines implementation. The ideal induction term is evaluated directly as the curl of
$\mathbf v\times\mathbf B$ in cylindrical coordinates. The resulting components of the radiative flux and their spatial derivatives are evaluated numerically on the radiation grid,
% The radiation-driven source is then obtained from 
% %
% \begin{equation}
% \mathbf F_0
% =
% -\frac{\sigma_{\rm T}}{e}
% \nabla\times\mathbf F_{\rm rad},
% \label{eq:F0_final}
% \end{equation}
% %
and the resulting two-dimensional source field is interpolated onto the magnetic-field evolution grid solving Equation \ref{eq:induction}. The radiation calculation is performed independently of the magnetic evolution, with no magnetic back-reaction
on the radiation field.
%Hall and resistive terms can be switched on
%or off independently.

Zero-Dirichlet boundary conditions are imposed on all three magnetic
field components at the radial and vertical outer boundaries,
\begin{equation}
B_r=B_\phi=B_z=0.
\end{equation}
The velocity and radiation fields are prescribed throughout the
calculation; consequently, the evolution is kinematic and does not include magnetic back-reaction on the accretion flow or radiation field.

We calculate the evolution of the radiation-generated magnetic field and
determine the time $t_{\rm RAD}$ required to reach the adopted equipartition
field as a function of coronal luminosity $l_c$, for the fiducial coronal
size $x_c=20$.
%The resulting time-dependent magnetic field is used to determine the radiation-driven magnetization time (time needed to amplify magnetic fields to dynamically significant values or equipartition with gas pressure) as a function of coronal luminosity $l_c$ and size $x_c$. 
These times are subsequently compared with the corresponding MRI amplification timescale ($t_{\rm MRI}$), allowing the radiation mechanism to be assessed as a competing channel for rapid magnetic-field generation and amplification.

\begin{table*}[t]
\centering
\caption{Comparison between magnetorotational instability (MRI) and the radiation-triggered magnetic-field generation mechanism.}
%\begin{tabular}{@{}p{3.8cm}p{5.2cm}p{6.6cm}@{}}
\begin{tabular}{lll}
\hline
\textbf{Property} & \textbf{MRI} & \textbf{Radiation-triggered mechanism} \\
\hline
Physical origin &
Magnetorotational instability &
Non-conservative radiation force ($\nabla\times\mathbf{F}_{\rm rad}\neq 0$) \\

Primary role &
Amplifies magnetic fields &
Generates and amplifies magnetic fields \\

Seed magnetic field &
Required &
Not required \\

Initial magnetic topology &
Depends on the assumed seed field &
Self-generated, ordered large-scale poloidal field \\

Growth law &
Exponential ($\propto e^{3\Omega t/4}$) &
Linear to quadratic \\

Characteristic growth rate &
Depends on location in the disc &
Depends on $l_c$, and $v_\phi$\\

Dominant physical ingredients &
Differential rotation + weak magnetic field &
Radiation + differential rotation\\

%Parameter dependence & Primarily local disc's differential rotation & Radiation luminosity, source geometry, and disc rotation \\

Dominant regime &
General ionized accretion discs &
Luminous, compact and rotating radiation sources \\
\hline
\end{tabular}
\label{tab:comparison}
\end{table*}
{\bf Radiation versus MRI.}
In Appendix \ref{app:analytic} we develop the analytical picture underlying the two
magnetization mechanisms. We compare the relative characteristics of the two underlying mechanisms in Table \ref{tab:comparison}.
Whereas the MRI amplifies a pre-existing seed field exponentially, the
radiation--shear mechanism produces quadratic growth through linear
radiation-driven field generation followed by rotational winding. The
corresponding timescales for the field to reach the adopted equipartition
strength $B_{\rm eq}$ are derived in Appendix B, Equations \ref{eq:trad_general} and \ref{eq:MRI_exp} respectively:
\begin{equation}
t_{\rm RAD}
\simeq
\left(
\frac{4B_{\rm eq}}
{3\Omega |K_r|}
\right)^{1/2},
\end{equation}
and
\begin{equation}
t_{\rm MRI}(x_{\rm MRI})=
\frac{19\sqrt{2}\pi r_g}{f_{\rm rot}c}
\sqrt{x_{\rm MRI}}\,(x_{\rm MRI}-1).
\label{eq:6}
\end{equation}
The radiation-driven timescale used in the parameter-space comparison is obtained directly from the numerical magnetic-field evolution. The characteristic time $t_{\rm MRI}$ is evaluated at location $x_{\rm MRI}$ (in units of $r_g$ and treated as a free parameter, see equation \ref{eq:MRI_exp}) while $t_{\rm RAD}$ is evaluated at the location where the radiation-driven calculation produces its maximum magnetic-field strength.
Here, $K_r$ is the growth rate of the radial magnetic field component
$B_r$ and depends on the coronal size, geometry, and luminosity; it is
therefore evaluated numerically. The MRI timescale is obtained by adopting
an amplification time of $1.5$ orbital periods for three decades of
field amplification and extrapolating linearly with
$\log_{10}(B_{\rm eq}/B_0)$. For the adopted
$B_0=10^{-12}\,\mathrm{G}$ and $B_{\rm eq}=10^7\,\mathrm{G}$, the required
amplification is 19 decades, giving $t_{\rm MRI}\simeq9.5\,t_{\rm orb}$.
In Equation \ref{eq:6}, 
%$x_{\rm MRI}=r_{\rm MRI}/r_g$, 
$f_{\rm rot}$ is the sub-Keplerian
rotation factor, taken as $f_{\rm rot}=0.5$ when MRI timescales are estimated inside the corona, and $f_{\rm rot}=1$  when obtained inside the Keplerian disc. The expression above follows from the adopted
Paczynski--Wiita velocity profile.
To compare the two underlying mechanisms, we define the timescale ratio
\begin{equation}
\mathcal{R}(l_c,x_{\rm MRI})
\equiv
\frac{t_{\rm RAD}(l_c)}
     {t_{\rm MRI}(x_{\rm MRI})}.
\label{eq:timescale_ratio}
\end{equation}
Thus for a parameter space comprising $l_c$ and $x_{\rm MRI}$, the dominance of the mechanisms is determined as
$$
\begin{aligned}
\mathcal{R}<1
&\quad\Rightarrow\quad
\text{radiation-driven magnetization is faster},\\
\mathcal{R}>1
&\quad\Rightarrow\quad
\text{MRI is faster}.
\end{aligned}
$$
The boundary $\mathcal{R}=1$ separates the two regimes in the
$(l_c,x_{\rm MRI})$ parameter space. %The radiation timescale is obtained from the numerical evolution for each $(l_c,x_{\rm MRI})$ configuration, while
%MRI is not explicitly modelled inside the corona because the present calculation treats the coronal flow as a prescribed kinematic structure with sub-Keplerian rotation rather than a self-consistently evolving MHD flow.
%Thus, each timescale is evaluated at the characteristic location of the corresponding magnetization process. %The resulting $\mathcal{R}(l_c,x_c)$ is presented in the next section. 

\section{Results}
\label{sec_results}

% \begin{figure}[t]
%     \centering
%     \includegraphics[width=\columnwidth]{B_over_Beq_vs_time.eps}
%     \caption{Time evolution of the maximum magnetic-field strength
%     normalized to the local equipartition field, $B/B_{\rm eq}$, for
%     different coronal luminosities $l_c$ at fixed coronal size
%     $x_c=50$. The horizontal line marks $B=B_{\rm eq}$.}
%     \label{fig:B_over_Beq_time}
% \end{figure} 

\begin{figure}[t]
    \centering
    \includegraphics[width=\columnwidth]{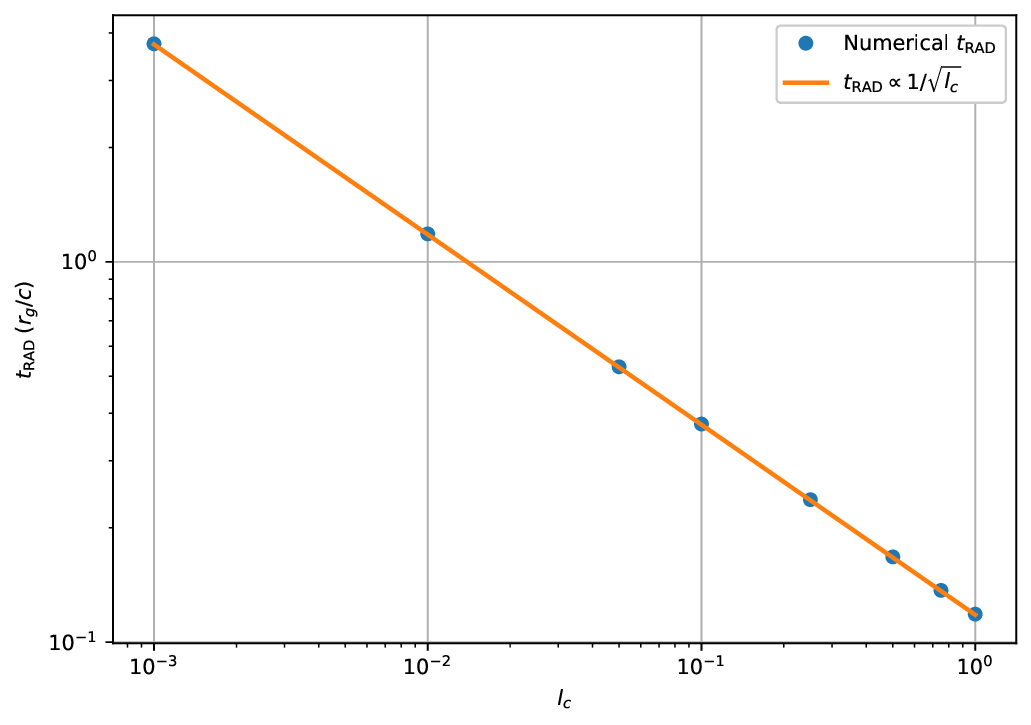}
    \caption{Radiation-driven magnetization timescale, $t_{\rm RAD}$, as a
function of coronal luminosity $l_c$. Coronal size
$x_c = 20$ assumed. The timescale shows the expected $t_{\rm RAD} \propto 1/\sqrt{l_c}$ behaviour. 
%The horizontal dashed lines indicate the corresponding MRI amplification timescales, ($t_{\rm MRI}$, independent of $l_c$). Radiation-driven magnetization is faster than MRI in the region to the right of the crossing between these lines, corresponding to a minimal coronal luminosity. 
}
    \label{fig:t_rad_vs_lc_xs}
\end{figure}

\begin{figure}[t]
    \centering
    \includegraphics[width=\columnwidth]{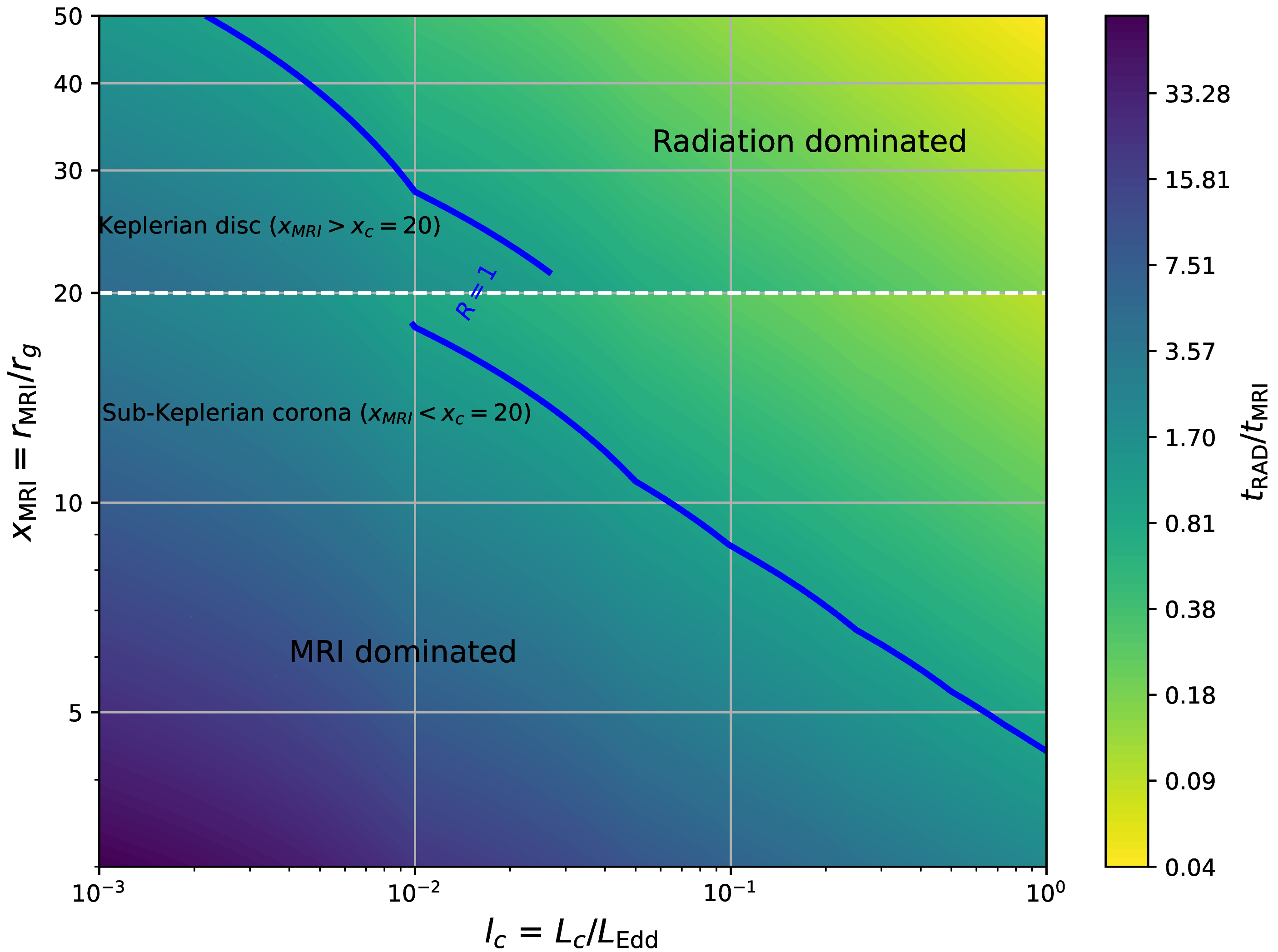}
    \caption{Ratio of the radiation-driven magnetization timescale to the
magnetorotational-instability (MRI) amplification timescale,
$t_{\rm RAD}/t_{\rm MRI}$, over the $(l_c,x_{\rm MRI})$ parameter space.
The blue curve denotes the boundary $t_{\rm RAD}/t_{\rm MRI}=1$, separating
the radiation-dominated regime, where radiation-driven magnetization reaches
the target field strength faster, from the MRI-dominated regime, where MRI
amplification is faster. The MRI timescale is evaluated at a given location $x_{\rm MRI}$ in the disc,
whereas $t_{\rm RAD}$ is determined from the time at which the maximum magnetic-field strength produced in the radiation calculation for the corresponding $l_c$ reaches $B_{\rm eq}$. The dashed horizontal line at $x_{\rm MRI}=x_c=20$ separates the outer Keplerian disc ($x_{\rm MRI}>20$, $f_{\rm rot}=1$) from the
inner sub-Keplerian corona ($x_{\rm MRI}<20$, $f_{\rm rot}=0.5$). The break in the $t_{\rm RAD}/t_{\rm MRI}=1$ boundary originates from the different values of angular velocities in the disc and the corona (different values of $f_{\rm rot}$) assumed.
}.
\label{fig:MRI_radiation_parameter_space}
\end{figure}

We first examine the spatial structure and evolution of the
radiation-generated magnetic field. The corona being luminous, radiation from the Keplerian disc is neglected in this letter, and the source of radiation that generates the magnetic field is limited to the corona itself. The MRI is evaluated at a specified radial location $x_{\rm MRI}$, which may lie within the sub-Keplerian corona ($x_{\rm MRI}<x_c$) or in the
outer Keplerian disc ($x_{\rm MRI}>x_c$).
Figure~\ref{fig:Bmag_Beq_evolution}
shows the distribution of the magnetic field produced by corona radiation and subsequently amplified. The magnetic-field strength is normalized to the characteristic equipartition field, $|\mathbf{B}|/B_{\rm eq}$ (Equation \ref{eq:Beq}), at three
representative times for a corona with $x_c=20 $, $l_c=1$ (1 Eddington luminosity). The height $z_m =3$ (in $r_g$)  is kept constant in this letter. As one can see from appendix \ref{app:corona_geometry}, the radiation field is independent of the corona size once $x_c \gg 1$ because the majority of the flux is produced in the inner coronal regions. 
%unless specified. 
The field develops over the region surrounding the corona and grows rapidly with time, reaching and locally exceeding the equipartition strength. This provides a direct view of the spatial development of the radiation-triggered magnetic field. % before considering its dependence on the coronal parameters.

% The time evolution of the radiation-generated magnetic
% field for different coronal luminosities is shown in Figure~\ref{fig:B_over_Beq_time}.
% Here we show the maximum field strength normalized to the local equipartition
% field, $B/B_{\rm eq}$, for $x_c=50$ and
% $l_c=0.1,\ 0.5,\ 0.01,$ and $0.001$. All calculations use
% $M=10M_\odot$ and the prescribed velocity profile with
% $f_{\rm rot}=0.5$. The magnetic field is initialized at zero, so the
% growth is entirely radiation-triggered. Increasing $l_c$ substantially
% reduces the time required to reach equipartition, consistent with the
% scaling $t_{\rm RAD}\propto l_c^{-1/2}$ derived above. For the more
% luminous coronae, the field reaches $B_{\rm eq}$ on the dynamical
% timescale of the inner accretion flow.

Figure~\ref{fig:t_rad_vs_lc_xs} shows the dependence of the radiation-driven magnetization timescale on coronal luminosity for parameters considered in Figure \ref{fig:Bmag_Beq_evolution}. For all configurations,
$t_{\rm RAD}$ decreases systematically with increasing $l_c$, approximately
following the expected $l_c^{-1/2}$ scaling as $K_r \propto l_c$ in Equation \ref{eq:trad_general}.% At a fixed luminosity, more compact coronae generally produce faster radiation-driven magnetization. Large corona implies slower magnetic field generation processes, in which the MRI becomes comparatively slower, implying that a dimmer corona is sufficient to produce dominant radiation-driven magentization. 

%We next compare the radiation-triggered magnetization timescale with
%the corresponding MRI amplification timescale. 
Figure~\ref{fig:MRI_radiation_parameter_space}
shows $\mathcal{R}(l_c,x_{\rm MRI})={t_{\rm RAD}}/{t_{\rm MRI}}$
over the $(l_c,x_{\rm MRI})$ parameter space, with $l_c$ spanning
$10^{-3}$--$1$ and $x_{\rm MRI}$ spanning $3$--$50$. Here,
$t_{\rm MRI}$ is evaluated at the location $x_{\rm MRI}$ and is treated
as a free parameter, allowing the MRI amplification time to be compared
at different locations in the accretion flow. The radiation timescale
$t_{\rm RAD}$, in contrast, is obtained from the radiation-driven
magnetic-field calculation for each $l_c$, as the time required for the
maximum magnetic field generated by the radiation field to reach
$B_{\rm eq}$. This maximum field is generally attained near the apex of
the corona, as illustrated in Figure~\ref{fig:Bmag_Beq_evolution}. The dashed
horizontal line at $x_{\rm MRI}=x_c=20$ separates the outer Keplerian
disc, where $f_{\rm rot}=1$, from the inner sub-Keplerian corona, where
$f_{\rm rot}=0.5$. The boundary $\mathcal{R}=1$ therefore separates the
radiation-dominated regime, where radiation-driven magnetization reaches
$B_{\rm eq}$ faster, from the MRI-dominated regime. Radiation-driven
magnetization dominates for sufficiently large $l_c$, while MRI
amplification is faster for low-luminosity coronae.
%It should be noted that $t_{\rm MRI}$ presented here should be regarded as an idealized lower bound on the timescale of field growth due to MRI, as it assumes the survival of the fastest-growing mode and rapid exponential amplification from $10^{-12}\,\mathrm{G}$ to equipartition, corresponding to an increase of approximately 17 orders of magnitude. However, since the field growth is expected to slow as the magnetic field approaches equipartition and eventually saturates, this is a crude assumption. %The actual MRI amplification timescale may therefore be longer by a factor of a few than the values adopted here.
The obtained MRI amplification time scales, $t_{\rm MRI}$ (Equation \ref{eq:MRI_exp}) are in line with the amplification timescales reported in MRI simulations \citep{1995ApJ...440..742H,1996ApJ...464..690H,1996ApJ...463..656S}, including simulations with zero-net-flux stratified vertical fields \citep{2024MNRAS.530.2778D}. %However, these simulations typically begin with relatively strong seed fields, only a few orders of magnitude below equipartition. The time required to amplify a much weaker seed field to equipartition may therefore be longer than the timescales inferred from the cited simulations, or evaluated by Equation \ref{eq:timescales}. These departures from ideal exponential growth, together with the relatively large seed fields conventionally adopted in simulations, imply that the actual MRI amplification timescale from a very weak seed field may be longer by a factor of a few than the values considered here.

Taken together, our results demonstrate that non-conservative radiation fields provide an independent and dynamically significant channel for magnetic-field generation in black-hole accretion flows. Radiation generates the initial poloidal field, which is subsequently amplified by differential rotation, allowing radiation-driven magnetization to become faster than MRI for sufficiently luminous and compact coronae. Even where MRI is formally faster at low coronal luminosities, $t_{\rm RAD}$ remains within the same order of magnitude as $t_{\rm MRI}$ across the parameter space considered, including $l_c \ll 1$. Thus, radiation-driven field generation should be regarded not merely as an alternative to MRI, but as an accompanying mechanism that can contribute substantially to magnetic-field generation in luminous accretion systems. More broadly, the mechanism requires only a non-conservative radiation field and differential plasma motion, suggesting its relevance to other luminous compact astrophysical systems.

\appendix
\section{Corona geometry and coronal emission}
\label{app:corona_geometry}
The radiation field is constructed from an inclined inner coronal
emitting surface. The radiation field produced from the Keplerian
disc is found to be sub-dominant by the coronal emission, and is 
therefore not considered here. 
The detailed radiation-field construction follows Vyas \& Pe'er 
\citep{2025ApJ...988L..59V, 2026ApJ..1006..120V}. We use
cylindrical coordinates $(r,\phi,z)$ and define the dimensionless radius
$x\equiv r/r_g$.

The coronal emitting surface is represented by a straight inclined surface
in the meridional $(r,z)$ plane, parameterized by $0\leq s\leq1$ via
\begin{equation}
\begin{array}{l}
x(s)=x_{\rm i}+(x_c-x_{\rm i})s, \\
z(s)=z_m+(z_l-z_m)s .
\end{array}
\label{eq:corona_surface}
\end{equation}
Here, $x_c$ is the outer radial extent of the corona, while $x_{\rm i}$,
$z_l$, and $z_m$ specify its inner radial coordinate and lower and upper
heights, respectively. We adopt $x_{\rm i}=0.1$, $z_l=0.001$, and
$z_m=3$ (normalized to $r_g$). Rotation of this
surface about the symmetry axis gives the axisymmetric coronal emitting
surface.

The dimensionless length element along the inclined surface is
$
J=
\left[(x_c-x_{\rm i})^2+(z_m-z_l)^2\right]^{1/2}.
$
%\label{eq:Jcor}
%\end{equation}
%
Thus, the differential physical area of the emitting surface is
\begin{equation}
dA=r_g^2\,x(s)\,J\,ds\,d\phi.
\label{eq:dAcor}
\end{equation}
%
% or, after integration over the azimuthal angle,
% %
% \begin{equation}
% dA_{\phi}=2\pi r_g^2\,x(s)\,J\,ds .
% \end{equation}

The coronal specific intensity is assumed to decrease with cylindrical
radius as
%
%\begin{equation}
$
I_{\rm cor}(x)=C_{0}\,x^{-3},
$
%\label{eq:Icor}
%\end{equation}
%
where $C_{0}$ is the normalization constant of the coronal specific
intensity. The choice $I_{\rm cor}\propto x^{-3}$ is motivated by the
radial dependence of the local dissipative flux in a Keplerian
accretion disc and is adopted here as a phenomenological prescription
for the coronal emission.
It is fixed by requiring that the total luminosity emitted by
the coronal surface is
%
%\begin{equation}
$
L_{\rm cor}=l_cL_{\rm Edd}.
$
%\label{eq:Lcor}
%\end{equation}
%
Since differential area element on the corona surface is $dA \propto x\,ds$, the
luminosity emitted by a surface element has the weighting
\begin{equation}
dL_{\rm cor}\propto C_{0}\,x^{-2}\,ds .
\end{equation}
Consequently, the normalization of the specific intensity is
\begin{equation}
C_{0}=
\frac{l_cL_{\rm Edd}}
{2\pi r_g^2J
\displaystyle\int_0^1 x(s)^{-2}\,ds},
\label{eq:d0c}
\end{equation}
with
\begin{equation}
\int_0^1x(s)^{-2}\,ds
=
\frac{1}{x_c-x_{\rm i}}
\left(
\frac{1}{x_{\rm i}}-\frac{1}{x_c}
\right) = {1 \over x_c x_i}.
\label{eq:corona_norm_integral}
\end{equation}

The normalized intensity $I_{\rm cor}$ is then used directly in
the surface integral for the radiation flux. In particular, the
contribution from each coronal surface element to the flux at a field
point is proportional to
\begin{equation}
d\mathbf{F}_{\rm cor}
\propto
I_{\rm cor}(x)
(\hat{\mathbf n}\cdot\hat{\mathbf l})
\frac{\hat{\mathbf l}}{R^2}\,dA ,
\label{eq:corona_flux_element}
\end{equation}
The outward unit normal to the emitting surface is
\begin{equation}
\hat{\mathbf n}=
\frac{(z_m-z_l)\,\hat{\mathbf r}
+(x_c-x_{\rm i})\,\hat{\mathbf z}}{J}.
\label{eq:corona_normal}
\end{equation}
$\hat{\mathbf l}$ is the photon propagation direction, and $R$ is the
distance between the emitting element and the field point. Surface
visibility, projection, and the Doppler correction associated with the
motion of the emitting surface are included in the numerical radiation
integral following Vyas \& Pe'er \citep{2025ApJ...988L..59V, 2026ApJ..1006..120V}.

For an extended corona, $x_c\gg x_i$ and $J\simeq x_c$, giving
\begin{equation}
C_0\simeq \frac{l_cL_{\rm Edd}x_i}{2\pi r_g^2},
\end{equation}
which is independent of $x_c$. Moreover, the explicit $J$ in $dA$
cancels the $1/J$ dependence of the surface projection, while
$dL_{\rm cor}\propto x^{-2}dx$ suppresses contributions from the outer
corona. Thus, the radiation fluxes approach an asymptotic value for
large $x_c$.

% The combination of the radial intensity profile and surface geometry gives
% %
% \begin{equation}
% dL_{\rm cor}\propto x^{-2}\,dx ,
% \label{eq:dLcor}
% \end{equation}
% %
% so that the emission is weighted toward the inner part of the coronal surface. The total coronal luminosity remains fixed at $l_cL_{\rm Edd}$
% as $x_c$ or $z_m$ is varied; changes in the radiation field therefore arise from the redistribution of this fixed luminosity over the changing
% geometry and from the corresponding geometric propagation factors.

%\appendix{A}
\section{Analytical Model}
\label{app:analytic}
We now develop a simple analytical model to identify the scaling of the radiation-driven magnetization timescale (time needed to amplify magnetic fields to dynamically significant values or equipartition with gas pressure) and to provide a direct comparison with the magnetorotational instability (MRI). We focus on the dominant sequence identified in the numerical calculations: radiation continuously generates a poloidal magnetic field, which is subsequently wound into a toroidal field by differential rotation. The Hall and resistive terms are retained in the numerical evolution, but are not required for the leading order scaling derived below.
We assume axisymmetry, ${\partial}/{\partial\phi}=0,$
and use the prescribed velocity field
$\mathbf{v} = (0,v_\phi,0),$ radially varying azimuthal velocity given in Section~\ref{sec:model}. In particular, the coronal flow is sub-Keplerian. 
The radiation source continuously generates a poloidal magnetic field. To leading order, the radial component can be written locally as
\begin{equation}
\frac{dB_r}{dt}
=
K_r,
\qquad
K_r\equiv F_{0,r},
\end{equation}
where $B_r$ is the radial magnetic-field component and $F_{0,r}$ is the radial component of the radiation-driven source defined by Eq.~(\ref{eq:radiation_source}). For a time interval over which the radiation field and the prescribed velocity field can be regarded as
fixed, this gives
\begin{equation}
B_r(t)
\simeq
K_r t,
\label{eq:Br_analytic}
\end{equation}
for the initial condition $B_r(0)=0$.

The generated radial field is then converted into an azimuthal field by
rotational shear. Neglecting, for the leading-order growth law, the direct
azimuthal radiation source and the sub-leading effects of Hall drift and
resistive diffusion, the shear contribution can be written as
\begin{equation}
\frac{dB_\phi}{dt}
\simeq
-q\Omega B_r,
\label{eq:Bphi_shear}
\end{equation}
where $B_\phi$ is the azimuthal magnetic-field component,
$\Omega\equiv v_\phi/r$ is the local angular velocity, and
\begin{equation}
q\equiv-\frac{d\ln\Omega}{d\ln r}
\end{equation}
is the dimensionless shear parameter. In the region
$3<x\leq x_c$, where $v_\phi=f_{\rm rot}v_{\rm K}$ with constant
$f_{\rm rot}=0.5$, $\Omega\propto r^{-3/2}$ and hence $q=3/2$, despite
the sub-Keplerian normalization of the rotation rate (For $x \gg 1$, the adopted Paczy\'nski--Wiita potential law approaches the Keplerian scaling 
$\Omega \propto r^{-3/2}$, so $q\simeq 3/2$.). Substitution of Eq.~(\ref{eq:Br_analytic}) gives
\begin{equation}
\frac{dB_\phi}{dt}
\simeq
-q\Omega K_r t,
\end{equation}
where $q$, $\Omega$, and $K_r$ are treated as locally constant over the
timescale considered. Integrating from $t=0$ gives
\begin{equation}
B_\phi(t)
\simeq
B_\phi(0)
-
\frac{1}{2}q\Omega K_r t^2.
\label{eq:Bphi_t2}
\end{equation}
For the initial condition $B_\phi(0)=0$,
\begin{equation}
|B_\phi(t)|
\simeq
\frac{1}{2}q\Omega |K_r|t^2
\propto t^2.
\end{equation}

This quadratic growth results from the coupled radiation--shear process:
radiation generates the poloidal field linearly in time, while differential
rotation continuously winds the generated field into the toroidal
component. The numerical evolution is used below to verify this
$t^2$ scaling and determine the corresponding magnetization timescale.

\subsection{Radiation-driven magnetization timescale}

We define the radiation-driven magnetization timescale, $t_{\rm RAD}$, as
the time required for the radiation-triggered magnetic field to reach the
local equipartition field $B_{\rm eq}$, defined by
$
{B_{\rm eq}^2}/{8\pi}\simeq P_{\rm gas}.
$
We adopt equipartition field following the estimate of
\citet{2017MNRAS.472L..20C},
\begin{equation}
B_{\rm eq}\simeq
10^8\,\dot{m}^{1/2}
\left(\frac{M}{M_\odot}\right)^{-1/2}\ {\rm G}.
\label{eq:Beq}
\end{equation}
For $M=10M_\odot$ and $\dot{m}=0.1$, this gives
\begin{equation}
B_{\rm eq}=10^7\ {\rm G}.
\end{equation}
This characteristic value is used as the target field strength for
both the radiation-driven and MRI amplification timescales.

Using Eq.~(\ref{eq:Bphi_t2}), the time required for the toroidal field
to reach the local equipartition field $B_{\rm eq}$ is
\begin{equation}
t_{\rm RAD}
\simeq
\left(
\frac{2B_{\rm eq}}
{q\Omega |K_r|}
\right)^{1/2},
\label{eq:trad_general}
\end{equation}
%
%where
%$q=-d\ln\Omega/d\ln r$ is the dimensionless shear parameter. 
% For the coronal region $3<x\leq x_c$, the adopted sub-Keplerian profile has
% $v_\phi=f_{\rm rot}v_{\rm K}$ with constant $f_{\rm rot}$. Thus
% $\Omega\propto r^{-3/2}$ and $q=3/2$, giving
% %
% \begin{equation}
% t_{\rm RAD}
% \simeq
% \left(
% \frac{4B_{\rm eq}}
% {3\Omega |K_r|}
% \right)^{1/2}.
% \label{eq:trad}
% \end{equation}
where
the radiation source is determined by the curl of the radiation flux,
\begin{equation}
K_r
=
-\frac{\sigma_{\rm T}}{e}
\left(
\nabla\times\mathbf{F}_{\rm rad}
\right)_r .
\label{eq:Kr}
\end{equation}

The calculation of the radiative time scale in Equation \ref{eq:trad_general} is consistent with the accurate numerical calculations, whose results are presented in Figure \ref{fig:t_rad_vs_lc_xs}, and provides a direct insight into its physical nature. 
%
% For the disc--corona geometries considered here, its characteristic
% magnitude scales approximately as $|K_r|
% \propto
% {l_c}/{x_c^2},
% \label{eq:K_scaling}
% $
% %
% where $x_c = {R_c}/{r_g}$
% %
% is the characteristic outer radius of the corona in units of $r_g$.
% At a fixed evaluation position, $B_{\rm eq}$ and $\Omega$ are fixed, so
% Eqs.~(\ref{eq:trad_general}) gives
% %
% $
% t_{\rm RAD}\propto l_c^{-1/2}.
% \label{eq:trad_l_scaling}
% $
% %
% Thus, increasing the coronal luminosity reduces the radiation-driven
% magnetization time approximately as the inverse square root of
% luminosity.

% When the characteristic evaluation radius scales with the coronal
% radius, $r\sim x_c r_g$, the local equipartition field must also be
% included in the size dependence. From Eq.~(\ref{eq:Beq}),
% $B_{\rm eq}\propto R^{-1}r^{-3/8}$, and hence
% $B_{\rm eq}\propto x_c^{-11/8}$ for $R\propto x_c$. Together with
% $\Omega_c\propto x_c^{-3/2}$ and
% $|K_r|\propto l_c x_c^{-2}$, Eq.~(\ref{eq:trad_general}) gives the approximate
% scaling
% %
% \begin{equation}
% t_{\rm RAD}
% \propto
% \left(
% \frac{B_{\rm eq}}
% {\Omega_c |K_r|}
% \right)^{1/2}
% \propto
% x_c^{17/16}l_c^{-1/2}.
% \label{eq:trad_rc}
% \end{equation}
% %
% This is only a local scaling estimate: the radiation source and
% $B_{\rm eq}(r,z)$ both vary across the full two-dimensional domain.
% The numerical calculations therefore determine the actual dependence
% of the magnetization timescale on $l_c$ and $x_c$.
%\subsection{MRI timescale}

\subsection{MRI field amplification}

For comparison with the radiation-driven amplification, we adopt a weak seed
field of $B_0=10^{-12}\,\mathrm{G}$ and a characteristic equipartition field
of $B_{\rm eq}=10^7\,\mathrm{G}$, following
\citet{2017MNRAS.472L..20C}. The required amplification is therefore
\begin{equation}
N=\log_{10}\left(\frac{B_{\rm eq}}{B_0}\right)=19
\end{equation}
decades. MRI channel-flow calculations show that equipartition can be reached in
$\simeq1.5$ orbital periods for an amplification of about three orders of
magnitude \citep{2010MNRAS.406..848L}. Assuming a linear scaling of the amplification
time with $\log_{10}(B_{\rm eq}/B_0)$, we obtain
\begin{equation}
t_{\rm MRI}\simeq
1.5\,t_{\rm orb}\frac{19}{3}
\simeq 9.5\,t_{\rm orb},
\end{equation}
where, for the adopted sub-Keplerian coronal rotation
$v_\phi=f_{\rm rot}v_K$ and
$v_K=c\sqrt{x/[2(x-1)^2]}$, the orbital time is
\begin{equation}
t_{\rm orb}(x)=
\frac{2\sqrt{2}\pi r_g}{f_{\rm rot}c}\sqrt{x}(x-1).
\end{equation}
Hence, the MRI amplification time as a function of radius
$x_{\rm MRI}=r/r_g$ is
\begin{equation}
t_{\rm MRI}(x_{\rm MRI})=
\frac{19\sqrt{2}\pi r_g}{f_{\rm rot}c}
\sqrt{x_{\rm MRI}}\,(x_{\rm MRI}-1). 
\label{eq:MRI_exp}
\end{equation}
%
% For the adopted $f_{\rm rot}=0.5$, this becomes
% %
% \begin{equation}
% \boxed{
% t_{\rm MRI}(x_{\rm MRI})=
% \frac{38\sqrt{2}\pi r_g}{c}
% \sqrt{x_{\rm MRI}}\,(x_{\rm MRI}-1)
% }.
% \end{equation}
This represents an extrapolation of the simulated amplification timescale
rather than a direct simulation from $10^{-12}$ to $10^7\,\mathrm{G}$.

% The comparison is based on the time required to reach the same magnetic
% field strength, rather than on the growth laws themselves: the
% radiation--shear mechanism grows approximately as $t^2$ from
% $B=0$, whereas MRI exponentially amplifies a pre-existing seed field.

% \begin{figure}[h!]
%     \centering
%     \includegraphics[width=9 cm, trim = 0 0 0 0, clip]{}
%     \caption{---
%     }
%     \label{lab_a1_a2_general}
% \end{figure}

\bibliography{ref1}% Produces the bibliography via BibTeX.

@ARTICLE{2010ApJ...716.1566A,
       author = {{Ando}, Masashi and {Doi}, Kentaro and {Susa}, Hajime},
        title = "{Generation of Seed Magnetic Field Around First Stars: Effects of Radiation Force}",
      journal = {Astrophysical Journal},
         year = 2010,
        month = jun,
       volume = {716},
       number = {2},
        pages = {1566-1572},
          doi = {10.1088/0004-637X/716/2/1566},
archivePrefix = {arXiv},
       eprint = {1005.1123},
 primaryClass = {astro-ph.CO},
       adsurl = {https://ui.adsabs.harvard.edu/abs/2010ApJ...716.1566A}
}

@ARTICLE{2014ApJ...782..108S,
       author = {{Shiromoto}, Yuki and {Susa}, Hajime and {Hosokawa}, Takashi},
        title = "{Generation of Magnetic Field on the Accretion Disk around a Proto-first-star}",
      journal = {Astrophysical Journal},
         year = 2014,
        month = feb,
       volume = {782},
       number = {2},
          eid = {108},
        pages = {108},
          doi = {10.1088/0004-637X/782/2/108},
archivePrefix = {arXiv},
       eprint = {1401.0905},
 primaryClass = {astro-ph.CO},
       adsurl = {https://ui.adsabs.harvard.edu/abs/2014ApJ...782..108S}
}

@ARTICLE{1977A&A....59..111B,
       author = {{Bisnovatyi-Kogan}, G.~S. and {Blinnikov}, S.~I.},
        title = "{Disk accretion onto a black hole at subcritical luminosity.}",
      journal = {Astronomy \& Astrophysics},
         year = 1977,
        month = jul,
       volume = {59},
        pages = {111-125},
       adsurl = {https://ui.adsabs.harvard.edu/abs/1977A&A....59..111B}
}

@ARTICLE{1998ApJ...508..859C,
       author = {{Contopoulos}, Ioannis and {Kazanas}, Demosthenes},
        title = "{A Cosmic Battery}",
      journal = {Astrophysical Journal},
         year = 1998,
        month = dec,
       volume = {508},
       number = {2},
        pages = {859-863},
          doi = {10.1086/306426},
archivePrefix = {arXiv},
       eprint = {astro-ph/9808223},
 primaryClass = {astro-ph},
       adsurl = {https://ui.adsabs.harvard.edu/abs/1998ApJ...508..859C}
}

@ARTICLE{2002ApJ...580..380B,
       author = {{Bisnovatyi-Kogan}, G.~S. and {Lovelace}, R.~V.~E. and {Belinski}, V.~A.},
        title = "{A Cosmic Battery Reconsidered}",
      journal = {Astrophysical Journal},
         year = 2002,
        month = nov,
       volume = {580},
       number = {1},
        pages = {380-388},
          doi = {10.1086/342876},
archivePrefix = {arXiv},
       eprint = {astro-ph/0207476},
 primaryClass = {astro-ph},
       adsurl = {https://ui.adsabs.harvard.edu/abs/2002ApJ...580..380B}
}

@ARTICLE{2006ApJ...652.1451C,
       author = {{Contopoulos}, Ioannis and {Kazanas}, Demosthenes and {Christodoulou}, Dimitris M.},
        title = "{The Cosmic Battery Revisited}",
      journal = {Astrophysical Journal},
         year = 2006,
        month = dec,
       volume = {652},
       number = {2},
        pages = {1451-1456},
          doi = {10.1086/507600},
archivePrefix = {arXiv},
       eprint = {astro-ph/0608701},
 primaryClass = {astro-ph},
       adsurl = {https://ui.adsabs.harvard.edu/abs/2006ApJ...652.1451C}
}

@ARTICLE{2008ApJ...674..388C,
       author = {{Christodoulou}, Dimitris M. and {Contopoulos}, Ioannis and {Kazanas}, Demosthenes},
        title = "{Simulations of the Poynting-Robertson Cosmic Battery in Resistive Accretion Disks}",
      journal = {Astrophysical Journal},
         year = 2008,
        month = feb,
       volume = {674},
       number = {1},
        pages = {388-407},
          doi = {10.1086/524699},
archivePrefix = {arXiv},
       eprint = {0706.3187},
 primaryClass = {astro-ph},
       adsurl = {https://ui.adsabs.harvard.edu/abs/2008ApJ...674..388C}
}

@ARTICLE{2023ApJS..264...32B,
       author = {{B{\'e}gu{\'e}}, D. and {Pe'er}, A. and {Zhang}, G. -Q. and {Zhang}, B. -B. and {Pevzner}, B.},
        title = "{cuHARM: A New GPU-accelerated GRMHD Code and Its Application to ADAF Disks}",
      journal = {Astrophysical Journal Supplement Series},
         year = 2023,
        month = feb,
       volume = {264},
       number = {2},
          eid = {32},
        pages = {32},
          doi = {10.3847/1538-4365/aca276},
archivePrefix = {arXiv},
       eprint = {2205.02484},
 primaryClass = {astro-ph.HE},
       adsurl = {https://ui.adsabs.harvard.edu/abs/2023ApJS..264...32B}
}

@ARTICLE{2021NewAR..9201610K,
       author = {{Komissarov}, Serguei and {Porth}, Oliver},
        title = "{Numerical simulations of jets}",
      journal = {New Astronomy Reviews},
         year = 2021,
        month = jun,
       volume = {92},
          eid = {101610},
        pages = {101610},
          doi = {10.1016/j.newar.2021.101610},
       adsurl = {https://ui.adsabs.harvard.edu/abs/2021NewAR..9201610K}
}

@ARTICLE{2022MNRAS.511.3795N,
       author = {{Narayan}, Ramesh and {Chael}, Andrew and {Chatterjee}, Koushik and {Ricarte}, Angelo and {Curd}, Brandon},
        title = "{Jets in magnetically arrested hot accretion flows: geometry, power, and black hole spin-down}",
      journal = {Monthly Notices of the Royal Astronomical Society},
         year = 2022,
        month = apr,
       volume = {511},
       number = {3},
        pages = {3795-3813},
          doi = {10.1093/mnras/stac285},
archivePrefix = {arXiv},
       eprint = {2108.12380},
 primaryClass = {astro-ph.HE},
       adsurl = {https://ui.adsabs.harvard.edu/abs/2022MNRAS.511.3795N}
}

@ARTICLE{1991ApJ...376..214B,
       author = {{Balbus}, Steven A. and {Hawley}, John F.},
        title = "{A Powerful Local Shear Instability in Weakly Magnetized Disks. I. Linear Analysis}",
      journal = {Astrophysical Journal},
         year = 1991,
        month = jul,
       volume = {376},
        pages = {214},
          doi = {10.1086/170270},
       adsurl = {https://ui.adsabs.harvard.edu/abs/1991ApJ...376..214B}
}

@ARTICLE{1995ApJ...440..742H,
       author = {{Hawley}, John F. and {Gammie}, Charles F. and {Balbus}, Steven A.},
        title = "{Local Three-dimensional Magnetohydrodynamic Simulations of Accretion Disks}",
      journal = {Astrophysical Journal},
         year = 1995,
        month = feb,
       volume = {440},
        pages = {742},
          doi = {10.1086/175311},
       adsurl = {https://ui.adsabs.harvard.edu/abs/1995ApJ...440..742H}
}

@ARTICLE{2011MNRAS.418L..79T,
       author = {{Tchekhovskoy}, Alexander and {Narayan}, Ramesh and {McKinney}, Jonathan C.},
        title = "{Efficient generation of jets from magnetically arrested accretion on a rapidly spinning black hole}",
      journal = {Monthly Notices of the Royal Astronomical Society},
         year = 2011,
        month = nov,
       volume = {418},
       number = {1},
        pages = {L79-L83},
          doi = {10.1111/j.1745-3933.2011.01147.x},
archivePrefix = {arXiv},
       eprint = {1108.0412},
 primaryClass = {astro-ph.HE},
       adsurl = {https://ui.adsabs.harvard.edu/abs/2011MNRAS.418L..79T}
}

@ARTICLE{2014ARA&A..52..529Y,
       author = {{Yuan}, Feng and {Narayan}, Ramesh},
        title = "{Hot Accretion Flows Around Black Holes}",
      journal = {Annual Review of Astronomy and Astrophysics},
         year = 2014,
        month = aug,
       volume = {52},
        pages = {529-588},
          doi = {10.1146/annurev-astro-082812-141003},
archivePrefix = {arXiv},
       eprint = {1401.0586},
 primaryClass = {astro-ph.HE},
       adsurl = {https://ui.adsabs.harvard.edu/abs/2014ARA&A..52..529Y}
}

@ARTICLE{1980A&A....88...23P,
       author = {{Paczy{\'n}sky}, B. and {Wiita}, P.~J.},
        title = "{Thick Accretion Disks and Supercritical Luminosities}",
      journal = {Astronomy \& Astrophysics},
         year = 1980,
        month = aug,
       volume = {88},
        pages = {23},
       adsurl = {https://ui.adsabs.harvard.edu/abs/1980A&A....88...23P}
}

@BOOK{1978mfge.book.....M,
       author = {{Moffatt}, H.~K.},
        title = "{Magnetic field generation in electrically conducting fluids}",
         year = 1978,
       adsurl = {https://ui.adsabs.harvard.edu/abs/1978mfge.book.....M}
}

@ARTICLE{2015ApJ...805..105C,
       author = {{Contopoulos}, Ioannis and {Nathanail}, Antonios and {Katsanikas}, Matthaios},
        title = "{The Cosmic Battery in Astrophysical Accretion Disks}",
      journal = {Astrophysical Journal},
         year = 2015,
        month = jun,
       volume = {805},
       number = {2},
          eid = {105},
        pages = {105},
          doi = {10.1088/0004-637X/805/2/105},
archivePrefix = {arXiv},
       eprint = {1501.05784},
 primaryClass = {astro-ph.HE},
       adsurl = {https://ui.adsabs.harvard.edu/abs/2015ApJ...805..105C}
}

@ARTICLE{2014ApJ...794...27K,
       author = {{Koutsantoniou}, Leela E. and {Contopoulos}, Ioannis},
        title = "{Accretion Disk Radiation Dynamics and the Cosmic Battery}",
      journal = {Astrophysical Journal},
         year = 2014,
        month = oct,
       volume = {794},
       number = {1},
          eid = {27},
        pages = {27},
          doi = {10.1088/0004-637X/794/1/27},
archivePrefix = {arXiv},
       eprint = {1405.6018},
 primaryClass = {astro-ph.HE},
       adsurl = {https://ui.adsabs.harvard.edu/abs/2014ApJ...794...27K}
}

@ARTICLE{2025ApJ...988L..59V,
       author = {{Vyas}, Mukesh Kumar and {Pe'er}, Asaf},
        title = "{Generation of Magnetic Fields around Black Hole Accretion Disks due to Nonconservative Radiation Fields}",
      journal = {Astrophysical Journal Letters},
         year = 2025,
        month = aug,
       volume = {988},
       number = {2},
          eid = {L59},
        pages = {L59},
          doi = {10.3847/2041-8213/aded11},
archivePrefix = {arXiv},
       eprint = {2505.10460},
 primaryClass = {astro-ph.HE},
       adsurl = {https://ui.adsabs.harvard.edu/abs/2025ApJ...988L..59V}
}

@ARTICLE{2020ARA&A..58..407D,
       author = {{Davis}, Shane W. and {Tchekhovskoy}, Alexander},
        title = "{Magnetohydrodynamics Simulations of Active Galactic Nucleus Disks and Jets}",
      journal = {Annual Review of Astronomy and Astrophysics},
         year = 2020,
        month = aug,
       volume = {58},
        pages = {407-439},
          doi = {10.1146/annurev-astro-081817-051905},
archivePrefix = {arXiv},
       eprint = {2101.08839},
 primaryClass = {astro-ph.HE},
       adsurl = {https://ui.adsabs.harvard.edu/abs/2020ARA&A..58..407D}
}

@ARTICLE{1977MNRAS.179..433B,
       author = {{Blandford}, R.~D. and {Znajek}, R.~L.},
        title = "{Electromagnetic extraction of energy from Kerr black holes.}",
      journal = {Monthly Notices of the Royal Astronomical Society},
         year = 1977,
        month = may,
       volume = {179},
        pages = {433-456},
          doi = {10.1093/mnras/179.3.433},
       adsurl = {https://ui.adsabs.harvard.edu/abs/1977MNRAS.179..433B}
}

@ARTICLE{2003PASJ...55L..69N,
       author = {{Narayan}, Ramesh and {Igumenshchev}, Igor V. and {Abramowicz}, Marek A.},
        title = "{Magnetically Arrested Disk: an Energetically Efficient Accretion Flow}",
      journal = {Publications of the Astronomical Society of Japan},
         year = 2003,
        month = dec,
       volume = {55},
        pages = {L69-L72},
          doi = {10.1093/pasj/55.6.L69},
archivePrefix = {arXiv},
       eprint = {astro-ph/0305029},
 primaryClass = {astro-ph},
       adsurl = {https://ui.adsabs.harvard.edu/abs/2003PASJ...55L..69N}
}

@ARTICLE{2019Galax...7...12C,
       author = {{Contopoulos}, Ioannis},
        title = "{Generation and Transport of Magnetic Flux in Accretion-Ejection Flows}",
      journal = {Galaxies},
         year = 2019,
        month = jan,
       volume = {7},
       number = {1},
          eid = {12},
        pages = {12},
          doi = {10.3390/galaxies7010012},
archivePrefix = {arXiv},
       eprint = {1901.01404},
 primaryClass = {astro-ph.HE},
       adsurl = {https://ui.adsabs.harvard.edu/abs/2019Galax...7...12C}
}

@ARTICLE{2025arXiv250815532W,
       author = {{Wallace}, John and {B{\'e}gu{\'e}}, Damien and {Pe'er}, Asaf},
        title = "{Direct Solution of the Time-Dependent Covariant Radiative Transfer Equation and its Coupling to General Relativistic Magnetohydrodynamics with cuHARM}",
      journal = {arXiv e-prints},
         year = 2025,
        month = aug,
          eid = {arXiv:2508.15532},
        pages = {arXiv:2508.15532},
          doi = {10.48550/arXiv.2508.15532},
archivePrefix = {arXiv},
       eprint = {2508.15532},
 primaryClass = {astro-ph.HE},
       adsurl = {https://ui.adsabs.harvard.edu/abs/2025arXiv250815532W}
}

@ARTICLE{1982MNRAS.199..883B,
       author = {{Blandford}, R.~D. and {Payne}, D.~G.},
        title = "{Hydromagnetic flows from accretion disks and the production of radio jets.}",
      journal = {Monthly Notices of the Royal Astronomical Society},
         year = 1982,
        month = jun,
       volume = {199},
        pages = {883-903},
          doi = {10.1093/mnras/199.4.883},
       adsurl = {https://ui.adsabs.harvard.edu/abs/1982MNRAS.199..883B}
}

@ARTICLE{2026ApJ..1006..120V,
       author = {{Vyas}, Mukesh Kumar and {Pe'er}, Asaf},
        title = "{Radiation-driven Origin of Dynamically Significant Magnetic Fields around Accretion Disks}",
      journal = {Astrophysical Journal},
         year = 2026,
        month = aug,
       volume = {1006},
       number = {2},
          eid = {120},
        pages = {120},
          doi = {10.3847/1538-4357/ae7f0c},
       adsurl = {https://ui.adsabs.harvard.edu/abs/2026ApJ..1006..120V}
}

@ARTICLE{1995ApJ...446..741B,
       author = {{Brandenburg}, Axel and {Nordlund}, Ake and {Stein}, Robert F. and {Torkelsson}, Ulf},
        title = "{Dynamo-generated Turbulence and Large-Scale Magnetic Fields in a Keplerian Shear Flow}",
      journal = {Astrophysical Journal},
         year = 1995,
        month = jun,
       volume = {446},
        pages = {741},
          doi = {10.1086/175831},
       adsurl = {https://ui.adsabs.harvard.edu/abs/1995ApJ...446..741B}
}

@ARTICLE{2019arXiv190409677J,
       author = {{Jafari}, Amir},
        title = "{Magnetic Fields in Accretion Disks: A Review}",
      journal = {arXiv e-prints},
         year = 2019,
        month = apr,
          eid = {arXiv:1904.09677},
        pages = {arXiv:1904.09677},
          doi = {10.48550/arXiv.1904.09677},
archivePrefix = {arXiv},
       eprint = {1904.09677},
 primaryClass = {astro-ph.HE},
       adsurl = {https://ui.adsabs.harvard.edu/abs/2019arXiv190409677J}
}

@ARTICLE{2003ARA&A..41..555B,
       author = {{Balbus}, Steven A.},
        title = "{Enhanced Angular Momentum Transport in Accretion Disks}",
      journal = {Annual Review of Astronomy and Astrophysics},
         year = 2003,
        month = jan,
       volume = {41},
        pages = {555-597},
          doi = {10.1146/annurev.astro.41.081401.155207},
archivePrefix = {arXiv},
       eprint = {astro-ph/0306208},
 primaryClass = {astro-ph},
       adsurl = {https://ui.adsabs.harvard.edu/abs/2003ARA&A..41..555B}
}

@article{event2021first,
  title={First M87 event horizon telescope results. VIII. Magnetic field structure near the event horizon},
  author={Event Horizon Telescope Collaboration and Akiyama, Kazunori and Algaba, Juan Carlos and Alberdi, Antxon and Alef, Walter and Anantua, Richard and Asada, Keiichi and Azulay, Rebecca and Baczko, Anne-Kathrin and Ball, David and others},
  journal={The Astrophysical Journal Letters},
  volume={910},
  number={1},
  pages={L13},
  year={2021},
  publisher={The American Astronomical Society}
}

@ARTICLE{1996ApJ...464..690H,
       author = {{Hawley}, John F. and {Gammie}, Charles F. and {Balbus}, Steven A.},
        title = "{Local Three-dimensional Simulations of an Accretion Disk Hydromagnetic Dynamo}",
      journal = {Astrophysical Journal},
         year = 1996,
        month = jun,
       volume = {464},
        pages = {690},
          doi = {10.1086/177356},
       adsurl = {https://ui.adsabs.harvard.edu/abs/1996ApJ...464..690H}
}

@ARTICLE{1996ApJ...463..656S,
       author = {{Stone}, James M. and {Hawley}, John F. and {Gammie}, Charles F. and {Balbus}, Steven A.},
        title = "{Three-dimensional Magnetohydrodynamical Simulations of Vertically Stratified Accretion Disks}",
      journal = {Astrophysical Journal},
         year = 1996,
        month = jun,
       volume = {463},
        pages = {656},
          doi = {10.1086/177280},
       adsurl = {https://ui.adsabs.harvard.edu/abs/1996ApJ...463..656S}
}

@ARTICLE{2024MNRAS.530.2778D,
       author = {{Dhang}, Prasun and {Bendre}, Abhijit B. and {Subramanian}, Kandaswamy},
        title = "{Shedding light on the MRI-driven dynamo in a stratified shearing box}",
      journal = {Monthly Notices of the Royal Astronomical Society},
         year = 2024,
        month = may,
       volume = {530},
       number = {3},
        pages = {2778-2794},
          doi = {10.1093/mnras/stae1011},
archivePrefix = {arXiv},
       eprint = {2308.07959},
 primaryClass = {astro-ph.HE},
       adsurl = {https://ui.adsabs.harvard.edu/abs/2024MNRAS.530.2778D}
}

@ARTICLE{1994MNRAS.267..235L,
       author = {{Lubow}, S.~H. and {Papaloizou}, J.~C.~B. and {Pringle}, J.~E.},
        title = "{Magnetic field dragging in accretion discs}",
      journal = {Monthly Notices of the Royal Astronomical Society},
         year = 1994,
        month = mar,
       volume = {267},
       number = {2},
        pages = {235-240},
          doi = {10.1093/mnras/267.2.235},
       adsurl = {https://ui.adsabs.harvard.edu/abs/1994MNRAS.267..235L}
}

@ARTICLE{2010ApJ...713...52D,
       author = {{Davis}, Shane W. and {Stone}, James M. and {Pessah}, Martin E.},
        title = "{Sustained Magnetorotational Turbulence in Local Simulations of Stratified Disks with Zero Net Magnetic Flux}",
      journal = {Astrophysical Journal},
         year = 2010,
        month = apr,
       volume = {713},
       number = {1},
        pages = {52-65},
          doi = {10.1088/0004-637X/713/1/52},
archivePrefix = {arXiv},
       eprint = {0909.1570},
 primaryClass = {astro-ph.HE},
       adsurl = {https://ui.adsabs.harvard.edu/abs/2010ApJ...713...52D}
}

@ARTICLE{2019ARA&A..57..467B,
       author = {{Blandford}, Roger and {Meier}, David and {Readhead}, Anthony},
        title = "{Relativistic Jets from Active Galactic Nuclei}",
      journal = {Annual Review of Astronomy and Astrophysics},
         year = 2019,
        month = aug,
       volume = {57},
        pages = {467-509},
          doi = {10.1146/annurev-astro-081817-051948},
archivePrefix = {arXiv},
       eprint = {1812.06025},
 primaryClass = {astro-ph.HE},
       adsurl = {https://ui.adsabs.harvard.edu/abs/2019ARA&A..57..467B}
}

@ARTICLE{2017MNRAS.472L..20C,
       author = {{Contopoulos}, I. and {Kazanas}, D. and {Fukumura}, K.},
        title = "{Magnetically advected winds}",
      journal = {Monthly Notices of the Royal Astronomical Society},
         year = 2017,
        month = nov,
       volume = {472},
       number = {1},
        pages = {L20-L24},
          doi = {10.1093/mnrasl/slx123},
archivePrefix = {arXiv},
       eprint = {1705.11026},
 primaryClass = {astro-ph.HE},
       adsurl = {https://ui.adsabs.harvard.edu/abs/2017MNRAS.472L..20C}
}

@ARTICLE{2010MNRAS.406..848L,
       author = {{Latter}, Henrik N. and {Fromang}, Sebastien and {Gressel}, Oliver},
        title = "{MRI channel flows in vertically stratified models of accretion discs}",
      journal = {Monthly Notices of the Royal Astronomical Society},
         year = 2010,
        month = aug,
       volume = {406},
       number = {2},
        pages = {848-862},
          doi = {10.1111/j.1365-2966.2010.16759.x},
archivePrefix = {arXiv},
       eprint = {1004.0109},
 primaryClass = {astro-ph.HE},
       adsurl = {https://ui.adsabs.harvard.edu/abs/2010MNRAS.406..848L}
}

@ARTICLE{2024ApJ...964L..25E,
       author = {{Event Horizon Telescope Collaboration} and {Akiyama}, Kazunori and {Alberdi}, Antxon and {Alef}, Walter and {Algaba}, Juan Carlos and others},
        title = "{First Sagittarius A* Event Horizon Telescope Results. VII. Polarization of the Ring}",
      journal = {Astrophysical Journal Letters},
         year = 2024,
        month = apr,
       volume = {964},
       number = {2},
          eid = {L25},
        pages = {L25},
          doi = {10.3847/2041-8213/ad2df0},
       adsurl = {https://ui.adsabs.harvard.edu/abs/2024ApJ...964L..25E}
}

\end{document}